\documentclass[aps,prl,reprint,groupedaddress,showpacs]{revtex4-1}

\usepackage{graphicx}
\begin{document}

% Use the \preprint command to place your local institutional report
% number in the upper righthand corner of the title page in preprint mode.
% Multiple \preprint commands are allowed.
% Use the 'preprintnumbers' class option to override journal defaults
% to display numbers if necessary
%\preprint{}

%Title of paper
\title{Learning in Neural Networks Based on a Generalized Fluctuation Theorem}

% repeat the \author .. \affiliation  etc. as needed
% \email, \thanks, \homepage, \altaffiliation all apply to the current
% author. Explanatory text should go in the []'s, actual e-mail
% address or url should go in the {}'s for \email and \homepage.
% Please use the appropriate macro foreach each type of information

% \affiliation command applies to all authors since the last
% \affiliation command. The \affiliation command should follow the
% other information
% \affiliation can be followed by \email, \homepage, \thanks as well.
\author{Takashi Hayakawa}
\email[]{takashi.hayakawa@riken.jp}
%\homepage[]{Your web page}
%\thanks{}
%\altaffiliation{}
\affiliation{RIKEN Brain Science Institute, \\ 2-1 Hirosawa, Wako, Saitama 351-0198, Japan}
\author{Toshio Aoyagi}
%\email[]{hayakawa@mbs.med.kyoto-u.ac.jp}
%\homepage[]{Your web page}
%\thanks{}
%\altaffiliation{}
\affiliation{Graduate School of Informatics, Kyoto University,\\ Yoshida-Honmachi, Sakyo-ku, Kyoto 606-8501, Japan}

%Collaboration name if desired (requires use of superscriptaddress
%option in \documentclass). \noaffiliation is required (may also be
%used with the \author command).
%\collaboration can be followed by \email, \homepage, \thanks as well.
%\collaboration{}
%\noaffiliation

\date{\today}

\begin{abstract}
Information maximization has been investigated as a possible mechanism of learning governing the self-organization that occurs within the neural systems of animals. Within the general context of models of neural systems bidirectionally interacting with environments, however, the role of information maximization remains to be elucidated. For bidirectionally interacting physical systems, universal laws describing the fluctuation they exhibit and the information they possess have recently been discovered. These laws are termed {\it fluctuation theorems}. In the present study, we formulate a theory of learning in neural networks bidirectionally interacting with environments based on the principle of information maximization. Our formulation begins with the introduction of a generalized fluctuation theorem, employing an interpretation appropriate for the present application, which differs from the original thermodynamic interpretation. We analytically and numerically demonstrate that the learning mechanism presented in our theory allows neural networks to efficiently explore their environments and optimally encode information about them.
% insert abstract here
\end{abstract}

% insert suggested PACS numbers in braces on next line
% insert suggested keywords - APS authors don't need to do this
%\keywords{}

%\maketitle must follow title, authors, abstract, \pacs, and \keywords
\maketitle

% body of paper here - Use proper section commands
% References should be done using the \cite, \ref, and \label commands
{\it Introduction} : The neural systems of animals are prominent as highly efficient systems for processing information concerning the external environment. Many authors have argued that the learning capability of neural systems is information-theoretically optimal by showing that several features of neural activity can be accounted for by positing information maximization (Infomax) in the learning of sensory signals and intrinsic dynamics in neural circuits \cite{Linsker:1988vn,Bell:1995vn,Bell:1997ve,Tanaka:2009bd,hayakawa2014biologically}. However, Infomax has not yet been investigated in a general context. In particular, although a real neural system interacts bidirectionally with its environment, not only receiving sensory signals and organizing intrinsic dynamics accordingly, but also generating motor outputs that influence the environment, Infomax learning has not been clearly formulated in this context (see Fig.1(a)). For example, the formulation of Infomax must be generalized in order to facilitate its application to the following type of model. One of the standard models of the interaction between neural systems and environments employs the Markov decision process. In this model, we consider discrete-time ($t\in \mathbf{Z}$) stationary Markovian dynamics of a stochastic neural network with $N$ binary-valued neurons $x^t\in \{ 0,1\} ^{N}$ interacting with an environment $y^t$ that takes values in a discrete state space $\mathcal{Y}$  \cite{Sallans:2004ul,Ackley:1985wk,smolensky1986parallel}. The neural elements $\{ x_i^{t}\} $ ($1\leq i\leq N$) receive inputs from the environment in such a manner that $x^t$ realizes values stochastically according to a conditional probability $\pi (x^t|y^t)$. This conditional probability depends on the model parameters, such as the synaptic strength, and it changes slowly during the learning process through the adjustment of the model parameters. Then, the state of the environment at the next timestep, $y^{t+1}$, is obtained stochastically with a transition probability $\mu (y^{t+1}|y^t,x^t)$ that is determined by the current states of the neural network and environment. 
\begin{figure}
\includegraphics[width=85mm]{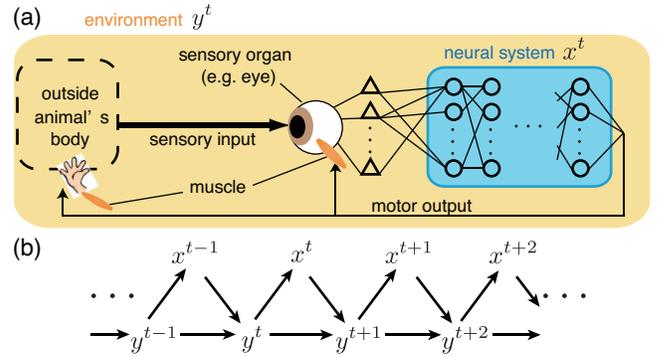}
\caption{(Color). (a) Interactions between a neural system and its environment, (b) Representation of the dynamics as a causal network.}
\end{figure}
\\
\\
For interacting physical systems, there are recently discovered universal laws called {\it fluctuation theorems} \cite{Jarzynski:313804,Crooks:377970,Evans:1993tm,Sagawa:2010df,Seifert:2012es,Ito:2013gp} that relate nonequilibrium physical quantities to informational quantities such as mutual information. In particular, the dynamics of the neural network and the environment mentioned above are represented in the form of a causal network with regard to which a generalized version of the fluctuation theorem has been investigated \cite{Ito:2013gp} (Fig.1(b)). Because fluctuation theorems describe informational quantities for interacting systems, it is natural to hypothesize that they may provide a description of a key aspect of the learning behavior exhibited by neural systems. Although information thermodynamic considerations have been investigated in the context of learning systems in a few pioneering studies \cite{Still:2012dfa,Still:2014it}, it has not been determined whether informational quantities are actually maximized in such systems in some systematic way. In this Letter, we study this question. 
\\
\\
We derive a novel type of Infomax learning, starting from the version of the integral fluctuation theorem presented in \cite{Ito:2013gp}, which provides the following inequality relating the average entropy production $\mathrm{E} [\sigma ]$ of the neural network and the transfer entropy $\mathrm{I} _{x\rightarrow y}$ from the neural network to its environment:
\begin{eqnarray}
\mathrm{I} _{x\rightarrow y}\geq -\mathrm{E} [\sigma ] , \ \ \ \sigma \equiv \log \frac{\pi (x^{t+1}|y^{t+1})}{\pi (x^t|y^{t+1})} \label{conv_IFT}. 
\end{eqnarray}
Throughout this article, the expectation value $\mathrm{E} $ is taken with respect to the stationary distribution $p_s$ of the dynamics unless otherwise noted. The transfer entropy is defined as the conditional mutual information \cite{Schreiber:2000wo}:
\begin{eqnarray}
\mathrm{I} _{x\rightarrow y}\equiv \mathrm{I} [y^{t+1};x^t|y^{t}]. \label{transfer_entropy}
\end{eqnarray}
As in information theory \cite{cover2012elements}, the (conditional) mutual information between two variables is defined as the change in the (conditional) entropy of one of the two variables owing to the inclusion of the other variable as a conditioning variable. Explicitly, we have $\mathrm{I} [y^{t+1};x^t|y^t]=\mathrm{H} [y^{t+1}|y^t]-\mathrm{H} [y^{t+1}|x^t,y^t]$. The (conditional) entropy is defined in terms of the stationary distribution as $\mathrm{H} [y^{t+1}|y^t]=-\mathrm{E} [\log p_s(y^{t+1}|y^t)]$.
The quantity $\mathrm{I} _{x\rightarrow y}$ represents the amount of information that the neural system possesses about the future state of the environment. Thus, from the point of view of Infomax, it is a reasonable hypothesis that maximizing this quantity is an effective learning mechanism. However, it is necessary for the calculation of $\mathrm{I} _{x\rightarrow y}$ to directly estimate the transition probability of the environment, $\mu $, and this estimation apparently cannot be carried out by the neural network itself. Its lower bound, $-\mathrm{E} [\sigma ]$, on the other hand, can be computed within the neural network, because $\sigma $ is determined by the transition probability of the neural network, $\pi $, alone. With these in mind, it is natural to conjecture that neural systems attempt to optimize their acquisition of information about the future by adjusting $\pi $ in such a manner to maximize the quantity $-\mathrm{E} [\sigma ]$. However, note that the equality in the relation, $\mathrm{I} _{x\rightarrow y}\geq -\mathrm{E} [\sigma ]$, is not generally realized, and thus the maximization of $-\mathrm{E} [\sigma ]$ does not necessarily imply the maximization of $\mathrm{I} _{x\rightarrow y}$. In the next, we show how consideration of a generalized entropy production, allows us to overcome this problem. 
% Put \label in argument of \section for cross-referencing
%\section{\label{}}
\\
\\
{\it Generalized Fluctuation Theorem} : 
We prove the following inequality below: 
\begin{eqnarray}
\mathrm{I} _{x\rightarrow y}\geq -\mathrm{E} [\Theta _{\nu }^{t}]. \label{jsu_transfer}
\end{eqnarray}
Here, we define the following generalized forms of the entropy production in terms of a conditional distribution $\nu $:
\begin{eqnarray}
\hspace{-0.5cm} \Theta  _{\nu }^t\equiv \log \frac{\pi (x^{t+1}|y^{t+1})}{\nu (x^t|x^{t+1},y^{t})}, \ \hspace{-0.05cm} \widehat{\Theta } _{\nu }^t\equiv \Theta _{\nu }^t\hspace{-0.04cm} +\log \frac{p_s(x^t|y^t)}{p_s(x^{t+1}|y^{t+1})}. \label{generalized_entropy_production}
\end{eqnarray} 
We can regard $\nu $ as representing physical quantities computed in the neural system on the basis of $x^t, y^t, y^{t+1}$ and adjusted through learning (see supplemental materials). First, we have the apparent identity
\begin{eqnarray}
&\lefteqn{\exp \left [\log p_s(x^t|y^{t})+\log \pi (x^{t+1}|y^{t+1})-\widehat{\Theta } _{\nu }^t\right ]} &\nonumber \\ 
&&=\exp \left [\log p_s(x^{t+1}|y^{t+1})+\log \nu (x^t|x^{t+1},y^t)\right ]. \label{apparent_identity}
\end{eqnarray}
Multiplying both sides by $p_s(y^{t+1},y^{t})$ and summing them over relevant random variables, we obtain a generalized form of the fluctuation theorem:
\begin{eqnarray}
\mathrm{E} \left [e^{-I_{tr}^{t+1}-\widehat{\Theta } _{\nu }^t}\right ]=1, \ \ I_{tr}^{t+1}\equiv \log \frac{p_s(y^{t+1}|x^{t},y^{t})}{p_s(y^{t+1}|y^{t})} .\label{equality_transfer}
\end{eqnarray}
Applying Jensen's inequality ($\exp \mathrm{E} [F(Z)]<\mathrm{E} [\exp F(Z)]$, which applies to any random variable $Z$ and any function $F$) to Eq.(\ref{equality_transfer}) gives
\begin{eqnarray}
\mathrm{E} \left [-I_{tr}^{t+1}-\widehat{\Theta } _{\nu }^t\right ]\leq 0. \label{after_Jensen}
\end{eqnarray}
Noting that $\mathrm{E} [\widehat{\Theta } _{\nu }^t]=\mathrm{E} [\Theta _{\nu }^t]$ and $\mathrm{I} _{x\rightarrow y}=\mathrm{E} [I _{tr}^{t+1}]$, we have the inequality in Eq.(\ref{jsu_transfer}).
It is found that, for fixed $\pi $, the right-hand side of Eq.(\ref{jsu_transfer}) is maximal if and only if 
\begin{eqnarray}
\nu (x^t|x^{t+1},y^t)=p_s(x^t|x^{t+1},y^{t}). \label{equality_without}
\end{eqnarray}
Furthermore, we can prove that the equality in Eq.(\ref{jsu_transfer}) holds if and only if, in addition to Eq.(\ref{equality_without}), the mutual information takes the maximal value for the fixed $\pi $ and hence satisfies $\mathrm{I} [x^t;y^t]=\mathrm{H} [y^t]$, under suitable conditions (see supplemental materials). 
If the neural network has sufficient capacity, it is expected that there is some optimal $\pi $ that maximizes both $\mathrm{I} _{x\rightarrow y}$ and $\mathrm{I} [x^t;y^t]$, simultaneously. In this case, the above analysis implies that the optimal $\pi $ is obtained by maximizing $-\mathrm{E}[\Theta _{\nu }^t]$ with respect to $\pi $ and $\nu $. In conclusion, we find that, for a neural network with a large capacity, the maximization of $-\mathrm{E}[\Theta _{\nu }^t]$ leads to the maximization of $\mathrm{I} _{x\rightarrow y}$ and $\mathrm{I} [x^t;y^t]$. Because the maximization of $\mathrm{I} [x^t;y^t]$ served as the definition of Infomax in previous studies \cite{Linsker:1988vn,Bell:1995vn,Bell:1997ve}, the maximization of $-\mathrm{E}[\Theta _{\nu }^t]$ provides a generalized Infomax. 
\\
\\
{\it Structures of Neural Networks} : To maximize $-\mathrm{E}[\Theta _{\nu }^t]$, the neural network must be able to adjust $\pi $ and $\nu $ to optimal conditional distributions through learning. For this purpose, in the remainder of this Letter, we parameterize $\pi (x^t|y^t)$ as
 \begin{eqnarray}
\hspace{-0.7cm} \pi (x^{t}|y^{t})&=&\prod _{\ell =1}^{L}\prod _{i=N_{\ell -1}+1}^{N_{\ell }}\hspace{-0.3cm} \pi _i(x_i^{t}|y^{t}, \{ x_k^{t}\} _{k=1}^{N_{\ell -1}}), \label{conditional_form} \\
\pi _i(x_i^{t}&=&1|y^{t},\{ x_k^{t}\} _{k=1}^{N_{\ell -1}})=g(e _i^t), \nonumber \\
e _i^t&=&\sum _{1\leq j\leq M_{\ell }}\rho _{ij}\{ g(\xi _{(\ell ),j}^{t})-\frac{1}{2} \} -h _0,\nonumber \\
\xi _{(\ell ),j}^{t}&=&\sum _{1\leq k\leq d}v_{jk}^{(\ell )}y_k^t+\sum _{1\leq k\leq N_{\ell -1}}w_{jk}^{(\ell )}x_k^t-h _j^{(\ell )}.\label{transition_model}
\end{eqnarray}
Here, $g(e_i^t)$ is the logistic function $(1+\tanh (e_i^t))/2$. Equation(\ref{transition_model}) describes the situation in which each neuron computes its own transition probability, $\pi _{i}$, through the intermediate units $\xi _{(\ell ),j}^t$, with the adjustable parameters $\rho _{ij}, v_{jk}^{(\ell )}, w_{jk}^{(\ell )}$ and $h _j^{(\ell )}$ and the constant parameter $h_0$. These parameters represent the synaptic strengths and intrinsic properties of the neurons and intermediate units. Note that we assume a layered structure of the system, as illustrated in Fig.2, in which neuron $x_i^t$ in layer $\ell $ receives an input $g(e_i^t)$ from the neurons in layers $1$ through $\ell -1$ and the environment through the $\ell $-th intermediate layer. It is known that an arbitrary continuous mapping of $\{y_{k}^t\} _{k=1}^d$ and $\{ x_k^t\} _{k=1}^{N_{\ell -1}}$ to $e _i^{t}$ can be approximated by the last two lines in Eq.(\ref{transition_model}) to arbitrary precision if the number of the intermediate units, $M_{\ell }$, is sufficiently large \cite{Funahashi:1989wi}. Thus, any conditional probability of the form given in Eq.(\ref{conditional_form}) can be represented in terms of $g(e _i^t)$, as in Eq.(\ref{transition_model}) . Increasing the number of layers of the neural network increases its capability to represent various conditional probabilities. We believe that the capability to represent a wide variety of conditional probabilities will allow for realization of the optimal $\pi $, and therefore such capability is necessary for our purposes. We model $\nu $ in the same way as $\pi $. Note that $\nu $ is not used for the realization of neural states. We consider the situation in which only the values of $\log \nu (x^t|x^{t+1},y^t)$ are calculated through some biological mechanisms based on the realized states, $x^t$, $x^{t+1}$ and $y^t$ (see supplemental materials for details regarding $\nu $). \\
\\
\begin{figure}
\includegraphics[width=85mm]{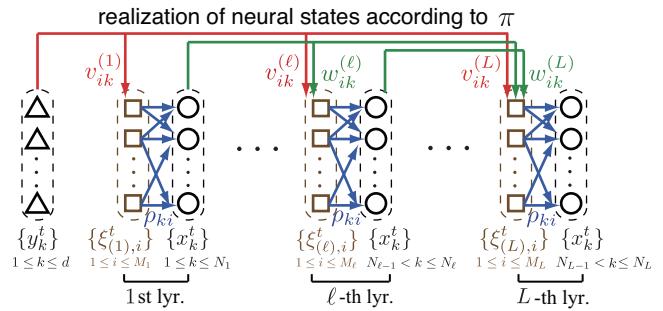}
\caption{(Color online). An illustration of the layered neural networks for the modeling of $\pi $ ($\nu $ is modeled in a similar manner).}
\end{figure}
{\it A Simple Model of Animals Learning to Explore Environments} : We have shown that neural systems can maximize the transfer entropy and mutual information through a learning mechanism based on a generalized fluctuation theorem. In order to characterize the present learning mechanism, we must clarify the role of the maximization of the transfer entropy in biological contexts, while that of the mutual information has been investigated in previous studies \cite{Linsker:1988vn,Bell:1995vn,Bell:1997ve}. In the following sections, we show that the maximization of the transfer entropy can be understood as a mechanism for the active exploration by an animal of its environment. In order to clearly demonstrate this effect in biological contexts, we introduce a learning problem in which an animal seeks to obtain rewards (e.g., food, water, etc.) through active exploration. \\
\\
Concretely, an animal with a neural system represented by the state $x^t$ moves around in a two-dimensional grid. At each position in the grid, a value of a reward associated with that position is defined (Fig.3(a)). Specifically, in each timestep, the animal takes either one step or zero steps, with the number and direction determined by the values of the specialized neurons, as shown in Fig.3(b). The state of the environment, $y^t$, is specified by the position of the animal and the status of reward configuration in the grid. At each timestep, the animal ``receives'' the reward $r^{t}=r(y^{t})$ at its present position. As shown in Fig.3(a), at most positions in the grid, the reward takes a negative value fixed throughout the simulation. Such a negative reward is interpreted as a punishment. The size of the punishment is minimal in the center of the grid and increases in each direction moving away from the center. At eight (fixed) positions in the outer region of the grid, there are positive rewards. The value of each is initially $R$. If such a positive reward is visited by the animal, the reward at the position is 0 for the subsequent 100 timesteps and then reset to $R$. The animal receives inputs from the environment as twelve real variables $\{ y_k^t\} \ (1\leq k\leq 12)$. The inputs $\{ y_k^t\} $ consist of the coordinate values of the animal's position in the grid $(k=1,2)$, the presence or absence of a reward $R$ at the animal's current position ($y_{3}^t=1\ \mathrm{or} \ 0$) and the values of the rewards at all positions within one step of the current position $(4\leq k\leq 12)$, as shown in Fig.3(c). This set of values allows the animal to predict the immediate consequence of its movement. Initially, the model parameters that determine $\pi $ are set in such a way that the animal primarily attempts to avoid negative rewards, mimicking the innate behavior of real animals (see supplemental materials). With this model, it is very natural to consider maximization of the average reward, $\mathrm{E} [r^t]$, by adjusting the animal's behavior represented by $\pi$, because animals must do so for survival. This maximization problem is called a {\it reinforcement learning problem}. However, in general, it is known that algorithms that simply maximize $\mathrm{E} [r^t]$ do not reach an optimal outcome in most realistic situations because there is a lack of new experiences, unless some mechanisms for active exploration are included \cite{Sutton:1998:IRL:551283,wiering2012reinforcement}. In the present case, in order to obtain the rewards $R$ to realize a larger $\mathrm{E} [r^t]$, the animal must possess a mechanism that allows it to explore the outer region and tolerate the punishment incurred there. In the following, we show that maximization of the transfer entropy in addition to the average reward provides this mechanism.
\\
\\
\begin{figure}
\includegraphics[width=85mm]{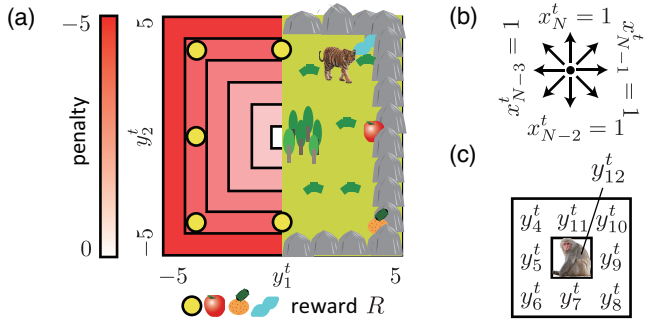}
\caption{(Color). A simple model of an animal exploring in a two-dimensional grid: (a) the configuration of reward in the grid, (b) animal's movement according to the values of the specialized neurons, (c) surrounding reward values as input variables.}
\end{figure}
We consider the following learning problem: 
\begin{eqnarray}
\max \ \left (\mathrm{I} _{x\rightarrow y}+\beta \mathrm{E} [r^t]\right ), \label{learning_problem}
\end{eqnarray}
where $\beta \geq 0$. First, we theoretically analyze the optimal $\pi $ for the above problem. Since the neural control over the environment is deterministic in the above model; that is, for given $x^t$ and $y^t$, we have $\mu (y^{t+1}|x^t,y^t)=1\ \mathrm{or} \ 0$, the optimization of $\pi (x^{t}|y^t)$ reduces to that of $\alpha (y^{t+1}|y^t)\equiv \sum _{x^{t}}\mu (y^{t+1}|x^t,y^{t})\pi (x^t|y^t)$. As we know from the basic theory of reinforcement learning, it is helpful for analysis of the maximization problem treated here to consider the following functions of $y\in \mathcal{Y}$: 
\begin{eqnarray}
\hspace{-0.6cm} V _{r, \alpha }^{(\gamma )}(y) \hspace{-0.02cm} &=&\hspace{-0.02cm} \mathrm{E} \hspace{-0.04cm} \left [\sum _{s=1}^{\infty }\gamma ^sr^{t+s}\Big{|} y^t=y\right ]\hspace{-0.06cm} -\hspace{-0.02cm} \mathrm{E} \hspace{-0.04cm} \left [\sum _{s=1}^{\infty }\gamma ^sr^{t+s}\right ]\hspace{-0.04cm} , \nonumber \\
\hspace{-0.6cm} V _{I, \alpha }^{(\gamma )}(y) \hspace{-0.02cm} &=&\hspace{-0.02cm} \mathrm{E} \hspace{-0.04cm} \left [\sum _{s=1}^{\infty }\gamma ^sI_{tr}^{t+s}\Big{|} y^t=y\right ]\hspace{-0.06cm} -\hspace{-0.02cm} \mathrm{E} \hspace{-0.04cm} \left [\sum _{s=1}^{\infty }\gamma ^sI_{tr}^{t+s}\right ]\hspace{-0.04cm} , \label{excess_function}
\end{eqnarray}
where $\gamma $ is a parameter satisfying $0\leq \gamma <1$. The above quantities with $\gamma \rightarrow 1$ represent the average amounts of ``excess reward'' and ``excess information'', obtained from the initial state $y$ until the system has relaxed into the steady state. This is analogous to the definition of the ``excess heat'' in steady-state thermodynamics \cite{Oono:1998uj,Sasa:2006gg}. With these limits, we can prove that the learning problem, Eq.(\ref{learning_problem}), has a unique optimal distribution $\alpha ^*$ of the following form (see supplemental materials):
\begin{eqnarray}
\hspace{-0.6cm} \alpha ^*\hspace{-0.04cm} (y^{t+1}|y^t)\hspace{-0.05cm} \propto \hspace{-0.05cm} \exp [\beta \{ r^{t+1}\hspace{-0.17cm}+\hspace{-0.06cm} V_{r,\alpha ^{*}}^{(1)}(y^{t+1})\hspace{-0.03cm} \} \hspace{-0.06cm} +\hspace{-0.06cm} V_{I,\alpha ^{*}}^{(1)}(y^{t+1})]. \label{Bellman_optimal_solution}
\end{eqnarray}
Inspecting Eq.(\ref{Bellman_optimal_solution}), we understand that the animal shows the following three types of behaviors determined by the value of $\beta $. In the case with finite $\beta (>0)$, the animal moves with high probability in a direction for which large future reward is expected, and with small (but non-zero) probability in a direction for which small future reward is expected. It is known that such exploratory behavior, with (infrequent) excursions in directions with low expected payoff, is necessary for neural systems to find larger rewards \cite{Sutton:1998:IRL:551283,wiering2012reinforcement}. Contrastingly, in the case with $\beta \rightarrow \infty $, the optimal behavior is deterministic, and exploration is stifled. In the case with $\beta =0$, the animal is completely insensitive to the values of reward. Hence, we see behavior that is a compromise between the drive to explore and the drive to acquire large rewards, represented by $\mathrm{I} _{x\rightarrow y}$ and $\mathrm{E} [r^t]$, respectively. 
\\
\\
{\it Numerical Simulations} : In order to confirm the theoretical results obtained in the above, we carried out simulations in which we maximized $\mathrm{E} [\beta r^t-\Theta _{\nu }^t]$ by applying a stochastic gradient algorithm to the model depicted in Fig.3 (see supplemental materials for the algorithms and discussion of its biological counterparts). It is expected that this maximization will result in the maximization in Eq.(\ref{learning_problem}). We first examine the case with $\beta =\infty $, i.e., that in which the animal attempts to maximize $\mathrm{E} [r^t]$ ($R=600$). In this case, we observe that the environmental entropy, $\mathrm{H} [y^t]$, decreases monotonically and that $\mathrm{E} [r^t]$ becomes fixed at zero (Fig.4(b),(c)). This indicates that the animal has learned only to avoid the outer areas and remains for all times at the origin. Hence, the learning has essentially failed. By contrast, setting $\beta =0.1$ and $R=0$, we observe that $-\mathrm{E} [\Theta _{\nu }^t]$ increases monotonically in Fig.4(a). We also observe that $\mathrm{H} [y^t]$ increases in a similar manner to $-\mathrm{E} [\Theta _{\nu }^t]$ and that $\mathrm{I} [x^t;y^t]$ almost realizes the maximal value, and satisfies $\mathrm{I} [x^t;y^t]=\mathrm{H} [y^t]$ (Fig.4(b)). Hence, we have confirmed that the maximization of $-\mathrm{E} [\Theta _{\nu }^t]$ leads to exploration and maximization of $\mathrm{I} [x^t;y^t]$, as theoretically predicted above. Finally, with $\beta =0.1$ and $R=600$, we find that the animal is able to increase $\mathrm{E} [r^t]$ through the exploration (Fig.4(c)).  
\begin{figure}
\includegraphics[width=85mm]{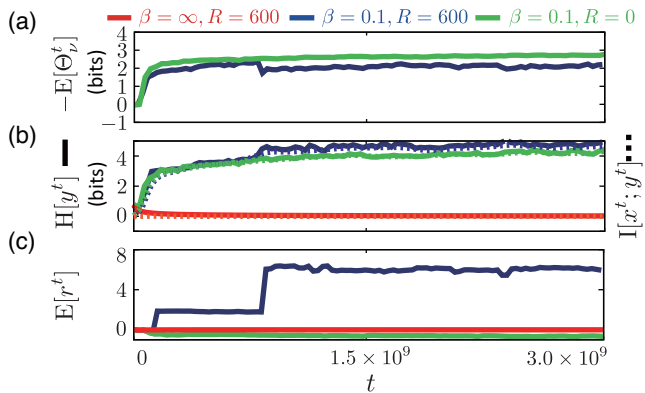}
\caption{(Color). The values of (a) $-\mathrm{E} [\Theta _{\nu }^t]$ (the red line is out of range), (b) $\mathrm{H} [y^t]$ and $\mathrm{I} [x^t;y^t]$, and (c) $\mathrm{E} [r^t]$, during the course of learning. Neural networks with $L=4$, $N_1=30$, $N_2=60$, $N_3=62$, $N_4=64$, $M_1=M_2=120$, and $M_3=M_4=60$ were simulated with $\beta=\infty ,0.1$, and $R=0,600$.}
\end{figure}
\\
\\
{\it Conclusion} : We have shown on the basis of theoretical and numerical analysis that assuming that the learning process exhibited by neural systems is based on a principle described by a generalized fluctuation theorem, this system will learn an effective form of exploring behavior (by maximizing the transfer entropy, $\mathrm{I} _{x\rightarrow y}$) and acquiring information about its environment (by maximizing the mutual information, $\mathrm{I} [x^t;y^t]$). Although informational quantities other than the transfer entropy have been considered as mechanisms for the exploration \cite{azar2012dynamic,peters2010relative,Still:2012dn,Ay:2012tu,Bialek:2001wv}, it has not been elucidated how those quantities are maximized in neural systems. We believe that use of the transfer entropy as a mechanism for exploration is more plausible, because the present learning mechanism can be utilized for it. Although the demonstration is limited to the case of Markovian environmental dynamics and neural networks without memory, this work will be generalized to more complex systems in the near future using the foundation laid by the present work.\\
\\
This work was supported by JST CREST from MEXT.
\newpage
\clearpage
\newpage
\begin{widetext}
\subsection{Supplemental Materials}
{\it Proof of The Maximization of The Mutual Information, $\mathrm{I} [x^t;y^t]$} : In this section, we prove that the maximization of the mutual information, $\mathrm{I} [x^t;y^t]$, and Eq.(\ref{equality_without}) are equivalent to equality in Eq.(\ref{jsu_transfer}), under suitable conditions. First, we note that we can replace $\nu (x^t|x^{t+1},y^t)$ by $\nu (x^t|x^{t+1},y^t,y^{t+1})$ in Eqs.(\ref{jsu_transfer}), (\ref{generalized_entropy_production}), (\ref{apparent_identity}), (\ref{equality_transfer}) and (\ref{after_Jensen}). In this case, equality in Eq.(\ref{jsu_transfer}) follows from equality in the Jensen inequality ($\exp F(Z)= \mathrm{E} [\exp F(Z)]$ with probability 1): 
\begin{eqnarray}
e^{-I_{tr}^{t+1}-\widehat{\Theta } _{\nu }^t}=\mathrm{E} [e^{-I_{tr}^{t+1}-\widehat{\Theta } _{\nu }^t}]=1.
\end{eqnarray}
This implies $-I_{tr}^{t+1}-\widehat{\Theta } _{\nu }^t=0$ with probability 1. By rearranging terms, we have
\begin{eqnarray}
\nu (x^t|x^{t+1},y^t,y^{t+1})&=&p_s(x^{t}|y^{t+1},y^{t},x^{t+1}). \label{equality_with}
\end{eqnarray}
Hence, by including $y^{t+1}$ as a conditioning variable of $\nu $, we can easily obtain the equality. However, by reducing the number of conditioning variables, we can also obtain the maximization of the mutual information, $\mathrm{I} [x^t;y^t]$, as we noted in the main text. We prove this in the following.  
\\
\\
First, we obtain an explicit expression of the optimal $\nu (x^t|x^{t+1},y^t)$ in Eq.(\ref{equality_without}) from the following inequality:
\begin{eqnarray}
\mathrm{E} \left [\log \frac{\nu (x^t|y^t,x^{t+1})}{p_s(x^t|y^t,x^{t+1})}\right ]=\sum _{y^t,x^{t+1}}p_s(y^t, x^{t+1})\sum _{x^t}p_s(x^t|y^t,x^{t+1})\log \frac{\nu (x^t|y^t,x^{t+1})}{p_s(x^t|y^t,x^{t+1})}\leq 0.
\end{eqnarray}
The above inequality is derived from the inequality $F-1\geq \log F$ for positive real $F$, and thus, the optimality condition in Eq.(\ref{equality_without}) is obtained from the equality, $F-1=\log F\leftrightarrow F=1$ with probability 1:
\begin{eqnarray}
\nu (x^t|x^{t+1},y^t)&=&p_s(x^{t}|y^{t},x^{t+1}).\label{equality_without_again}
\end{eqnarray}
Then, in order to analyze the equality condition of Eq.(\ref{jsu_transfer}) for $\nu (x^t|x^{t+1},y^t)$, we calculate the difference $\Delta $ between the values of $-\mathrm{E} [\Theta _{\nu }^t]$ as calculated with Eqs. (\ref{equality_without_again}) and (\ref{equality_with}), writing 
\begin{eqnarray}
\Delta =\mathrm{E} \left [\log \frac{p_s(x^t|y^{t},y^{t+1},x^{t+1})}{p_s(x^t|y^{t},x^{t+1})} \right ]=\mathrm{I} [y^{t+1};x^{t}|x^{t+1},y^{t}].
\end{eqnarray}
Because $-\mathrm{E} [\Theta _{\nu }^t]=\mathrm{I} _{x\rightarrow y}$ in the case with Eq.(\ref{equality_with}), we obtain
\begin{eqnarray}
\mathrm{I} _{x\rightarrow y}+\mathrm{E} [\Theta _{\nu }^{t}]=\mathrm{I} [y^{t+1};x^t|x^{t+1},y^{t}], \label{error_from_equality}
\end{eqnarray}
in the case with Eq.(\ref{equality_without_again}).
\\
\\
Next, we show that the condition 
\begin{eqnarray}
\mathrm{I} [y^{t+1};x^t|x^{t+1},y^{t}]=0, \label{control_condition_equality}
\end{eqnarray}
is equivalent to the maximization of the mutual information,
\begin{eqnarray}
\mathrm{I} [x^t;y^t]=\mathrm{H} [y^t], \label{maximal_mutual_information}
\end{eqnarray} 
on the assumption that the environmental state space $\mathcal{Y} $ is not too finely partitioned in comparison with the precision of neural control over the environment. Precisely, we assume that there is no coarse-grained partition $\mathcal{Y^{\prime }}$ of $\mathcal{Y}$ such that the neural control has the same precision on the two partitions $\mathcal{Y} $ and $\mathcal{Y} ^{\prime }$ of the environmental state space. We also assume that $p_s(y^{t+1}|y^t)\neq 1$ for all $y^t$ and $y^{t+1}$, which holds for most sets of values of the model parameters in a general model.
A coarse-grained partition $\mathcal{Y} ^{\prime }$ is a set of subcollections of $\mathcal{Y} $ such that $y^{\prime }\cap y^{\prime \prime }=\emptyset $ for any $y^{\prime }\neq y ^{\prime \prime }\in \mathcal{Y} ^{\prime }$ and $\cup _{y^{\prime }\in \mathcal{Y} ^{\prime }} y^{\prime }=\mathcal{Y} $.
For any such coarse-grained partition $\mathcal{Y} ^{\prime }$, we require
\begin{eqnarray}
\hspace{-0.8cm} \mathrm{I} [y^{t+1,\prime };x^t|y^t]\neq \mathrm{I} [y^{t+1};x^t|y^t],\ y^{t+1}\in y^{t+1,\prime }\in \mathcal{Y} ^{\prime }, \label{rough_partition}
\end{eqnarray}
where we have defined the random variable $y^{t+1,\prime }$, which takes values in $\mathcal{Y} ^{\prime }$ with
$y^{t+1}\in y^{t+1,\prime }$.
Under this assumption, we first show that the conditional mutual information, 
\begin{eqnarray}
\mathrm{I} [y^{t+1};x^{t+1}|y^{t}]=\sum _{x^{t+1},y^{t+1},y^t}p_s(y^t,y^{t+1},x^{t+1})\log \frac{p_s(y^{t+1}|x^{t+1},y^t)}{p_s(y^{t+1}|y^t)},
\end{eqnarray} 
must be maximal. Note that the conditional mutual information takes its maximal value and hence satisfies
\begin{eqnarray}
\mathrm{I} [x^{t+1};y^{t+1}|y^t]=\mathrm{H} [y^{t+1}|y^t] \label{maximality_conditional_mi}
\end{eqnarray}
if and only if $p_s(y^{t+1}|x^{t+1},y^t)=1$ with probability 1.
Thus, to obtain the desired result, we show that $p_s(y^{t+1}|x^{t+1},y^t)>0$ for multiple $y^{t+1}$ with some $y^t=y_0$ and $x^{t+1}=x_0$ contradicts Eq.(\ref{control_condition_equality}). \\
\\
First, we define the set
\begin{eqnarray}
\overline{y} =\{ y^{t+1}\in \mathcal{Y} |p_s(y^{t+1}|x^{t+1}=x_0,y^t=y_0)>0\} ,
\end{eqnarray}
and a coarse-grained partition of $\mathcal{Y} $ as 
\begin{eqnarray}
\mathcal{Y} ^{\prime }=\{ \{ y\} \} _{y\in {\mathcal{Y} \setminus \overline{y} }}\cup \{ \overline{y} \} .
\end{eqnarray}
Here, $\mathcal{Y} \setminus \overline{y}$ is the relative complement of $\overline{y}$ in $\mathcal{Y}$, which consists of all the elements of $\mathcal{Y}$ that are not contained in $\overline{y}$.
The assumption in Eq.(\ref{rough_partition}) requires 
\begin{eqnarray}
\mathrm{I} [y^{t+1};x^t|y^t]-\mathrm{I} [y^{t+1,\prime };x^t|y^t]&=&\mathrm{I} [y^{t+1},y^{t+1,\prime };x^t|y^t]-\mathrm{I} [y^{t+1,\prime };x^t|y^t] \nonumber \\
&=&\mathrm{I} [y^{t+1};x^t|y^{t+1,\prime },y^t]\nonumber \\
&>&0, \label{positive_control_assumption}
\end{eqnarray}
where the first equality holds because $y^{t+1}$ uniquely determines $y^{t+1,\prime }$ and thus the additional inclusion of $y^{t+1,\prime }$ in the first term does not affect the value of the conditional mutual information. Also, Eq.(\ref{control_condition_equality}) implies 
\begin{eqnarray}
\mathrm{I} [y^{t+1};x^t|y^t,x^{t+1}]=\mathrm{I} [y^{t+1};x^t|y^{t+1,\prime },y^t,x^{t+1}]=0.
\end{eqnarray}
Now, recall that the inclusion of additional conditioning variables (in this case, $y^{t+1,\prime }$) always reduces the value of the conditional mutual information. The right-hand side of the above equation can be written as
\begin{eqnarray}
&&\mathrm{E} \left [\log \frac{p_s(y^{t+1}|x^t,y^{t+1,\prime },y^t,x^{t+1})}{p_s(y^{t+1}|y^{t+1,\prime },y^t,x^{t+1})} \right ]\nonumber \\
&=&\mathrm{E} \left [\log \left \{ \frac{p_s(x^{t+1}|y^{t+1})p_s(y^{t+1}|y^{t+1,\prime },y^t,x^t)}{\sum _{\widetilde{y} ^{t+1}\in \mathcal{Y} }p_s(x^{t+1}|\widetilde{y} ^{t+1})p_s(\widetilde{y} ^{t+1}|y^{t+1,\prime },y^t,x^t)} \frac{\sum _{\widehat{y} ^{t+1}\in \mathcal{Y} }p_s(x^{t+1}|\widehat{y} ^{t+1})p_s(\widehat{y} ^{t+1}|y^{t+1,\prime },y^t)}{p_s(x^{t+1}|y^{t+1})p_s(y^{t+1}|y^{t+1,\prime },y^t)} \right \} \right ] \nonumber \\
&=&0, \label{contradiction_equality}
\end{eqnarray}
with the dummy variables $\widetilde{y} ^{t+1}$ and $\widehat{y} ^{t+1}$ having the same (conditional) distributions as $y^{t+1}$.
The above equality requires that the argument of the logarithm be 1 with probability 1, since $F-1\geq \log F$, $F-1= \log F\leftrightarrow F=1$ and 
\begin{eqnarray}
&&-\mathrm{E} \left [\log \frac{p_s(y^{t+1}|x^t,y^{t+1,\prime },y^t,x^{t+1})}{p_s(y^{t+1}|y^{t+1,\prime },y^t,x^{t+1})} \right ]\nonumber \\
&=&\sum _{x^t,y^{t+1,\prime },y^t,x^{t+1}}p_s(x^t,y^{t+1,\prime },y^t,x^{t+1})\sum _{y^{t+1}}p_s(y^{t+1}|x^t,y^{t+1,\prime },y^t,x^{t+1})\log \frac{p_s(y^{t+1}|y^{t+1,\prime },y^t,x^{t+1})}{p_s(y^{t+1}|x^t,y^{t+1,\prime },y^t,x^{t+1})} \nonumber \\
&\leq &\sum _{x^t,y^{t+1,\prime },y^t,x^{t+1}}p_s(x^t,y^{t+1,\prime },y^t,x^{t+1})\sum _{y^{t+1}}p_s(y^{t+1}|x^t,y^{t+1,\prime },y^t,x^{t+1})\left (\frac{p_s(y^{t+1}|y^{t+1,\prime },y^t,x^{t+1})}{p_s(y^{t+1}|x^t,y^{t+1,\prime },y^t,x^{t+1})}
-1\right ) \nonumber  \\
&=&0.
\end{eqnarray}
Hence, noting that $p_s(x^{t+1}=x_0|y^{t+1})>0$ for $y^{t+1}\in \overline{y} $, we have  
\begin{eqnarray}
\frac{p_s(y^{t+1}|y^{t+1,\prime }=\overline{y} ,y^{t},x^t)}{p_s(y^{t+1}|y^{t+1,\prime }=\overline{y} ,y^{t})} =\frac{\sum _{\widehat{y} ^{t+1}\in \mathcal{Y} }p_s(x^{t+1}=x_0|\widehat{y} ^{t+1})p_s(\widehat{y} ^{t+1}|y^{t+1,\prime }=\overline{y} ,y^t)}{\sum _{\widetilde{y} ^{t+1}\in \mathcal{Y} }p_s(x^{t+1}=x_0|\widetilde{y} ^{t+1})p_s(\widetilde{y} ^{t+1}|y^{t+1,\prime }=\overline{y} ,y^t,x^t)}, \  \ \ \forall y^{t+1}\in \overline{y} . \label{first_equation_for_contradiction}
\end{eqnarray}
Furthermore, Eq.(\ref{first_equation_for_contradiction}) with Eq.(\ref{positive_control_assumption}) implies 
\begin{eqnarray}
\frac{p_s(y^{t+1}|y^{t+1,\prime }=\overline{y} ,y^t,x^t)}{p_s(y^{t+1}|y^{t+1,\prime }=\overline{y} ,y^{t})}=c\neq 1, \label{cneq1}
\end{eqnarray}
for all $y^{t+1}\in \overline{y} $ and some $y^{t}$ and $x^t$, noting 
\begin{eqnarray}
\frac{p_s(y^{t+1}|y^{t+1,\prime },y^t,x^t)}{p_s(y^{t+1}|y^{t+1,\prime },y^{t})}=1, \ \ \ \forall y^{t+1,\prime }\in \mathcal{Y} ^{\prime } \setminus \{ \overline{y} \} . \label{always1}
\end{eqnarray}
Here, note that $c=1$ in Eq.(\ref{cneq1}) with Eq.(\ref{always1}) implies $\mathrm{I} [y^{t+1};x^t|y^{t+1,\prime },y^t]=0$, violating the assumption in Eq.(\ref{positive_control_assumption}). 
However, this implies 
\begin{eqnarray}
1=\sum _{y^{t+1}\in \overline{y} }p_s(y^{t+1}|y^{t+1,\prime }=\overline{y} ,y^t,x^t)=c\sum _{y^{t+1}\in \overline{y} }p_s(y^{t+1}|y^{t+1,\prime }=\overline{y} ,y^{t})=c\neq 1.
\end{eqnarray}
This contradiction completes the proof of the maximality of the conditional mutual information, Eq.(\ref{maximality_conditional_mi}). \\
\\
Next, we show the equivalence of the maximality of the conditional mutual information and the maximality of the mutual information. As we have discussed, Eq.(\ref{maximality_conditional_mi}) implies 
\begin{eqnarray}
p_s(y^{t+1}|x^{t+1},y^{t})=\frac{p_s(y^t|y^{t+1})p_s(y^{t+1}|x^{t+1})}{\sum _{\widetilde{y} ^{t+1}\in \mathcal{Y} }p_s(y^t|\widetilde{y} ^{t+1})p_s(\widetilde{y} ^{t+1}|x^{t+1})}=1,  \label{remove_cond}
\end{eqnarray}
with probability 1. Then, the assumption $p_s(y^{t+1}|y^{t})\neq 1$ implies that $p_s(y^t|y^{t+1})$ is positive for multiple $y^{t+1}$. Thus, the condition Eq.(\ref{remove_cond}) implies $p_s(y^{t+1}|x^{t+1})=1$ with probability 1, or equivalently, that the mutual information, $\mathrm{I} [x^{t};y^{t}]$, is maximal and hence satisfies 
\begin{eqnarray}
\mathrm{I} [x^t;y^t]=\mathrm{H} [y^{t}]. \label{max_mi}
\end{eqnarray}
Conversely, the equality in Eq.(\ref{max_mi})
implies 
\begin{eqnarray}
\mathrm{I} [y^{t+1};x^t|x^{t+1},y^t]&=&\mathrm{H} [y^{t+1}|x^{t+1},y^t]-\mathrm{H} [y^{t+1}|x^t,x^{t+1},y^t]\nonumber \\
&\leq &\mathrm{H} [y^{t+1}|x^{t+1}]\nonumber \\
&=&\mathrm{H} [y^{t}]-\mathrm{I} [x^t;y^t] \nonumber \\
&=&0.
\end{eqnarray}
Thus we have recovered the condition Eq.(\ref{control_condition_equality}). This completes the proof that equality in Eq.(\ref{jsu_transfer}) is equivalent to Eq.(\ref{equality_without}) and Eq.(\ref{maximal_mutual_information}).\\
\\
In the above proof, we note that different definitions of $\nu $ also lead to the maximization of the mutual information $\mathrm{I} [x^t;y^t]$ in different manners, although we do not note this point in the main text for the simplicity of presentation. Concretely, we consider the model described by the causal network in Fig.5 by splitting $x^t$ into $x_{(1)}^t$ and $x_{(2)}^t$, and define a generalized entropy production as
\begin{eqnarray}
\Theta _{\nu }^t=\log \frac{\pi (x_{(1)}^{t+1}|y^{t+1})}{\nu (x_{(1)}^t|x_{(2)}^t,y^{t+1})} .
\end{eqnarray} 
Then, in the same manner as above, we can prove that the equality, $-\mathrm{E} [\Theta _{\nu }^t]=\mathrm{I} _{x_{(1)}\rightarrow y}$, implies the maximization of the mutual information, $\mathrm{I} [x_{(2)}^t;y^{t,\prime }]=\mathrm{H} [y^{t,\prime }]$, in some coarse-grained partition $\mathcal{Y} ^{\prime }$ that satisfies
\begin{eqnarray}
\mathrm{I} [x_{(1)}^t;y^{t,\prime }|y^{t+1}]=\mathrm{I} [x_{(1)}^t;y^{t}|y^{t+1}].
\end{eqnarray} 
Further results in this direction will be investigated in the future reports.
\begin{figure}
\includegraphics[width=85mm]{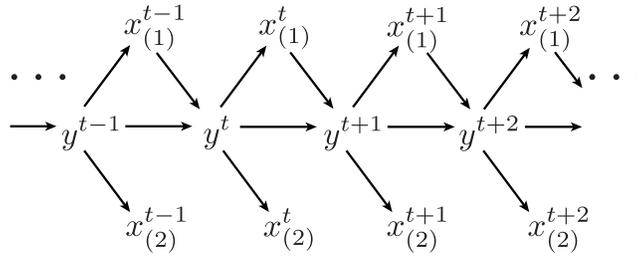}
\caption{Representation of the dynamics with an additional neural network $z^t$ as a causal network.}
\end{figure}
\\
\\
{\it Modeling of $\nu $ with a Neural Network} : We compute $\nu$ in the same way as $\pi $, explicitly writing 
\begin{eqnarray}
\nu (x^{t}|x^{t+1},y^t)&=&\prod _{\ell =1}^L\prod _{i=N_{\ell -1}+1}^{N_{\ell }}\nu _i(x_i^{t}|\{x_k^t\} _{k=1}^{N_{\ell -1}},x^{t+1},y^t), \nonumber \\
\nu _i(x_i^{t}&=&1|\{x_k^t\} _{k=1}^{N_{\ell -1}},x^{t+1}, y^t)=g(f _i^{t+1}), \nonumber \\
f _i^{t+1}&=&\sum _{1\leq j\leq M_{\ell }}\kappa _{ij}\{ g(\eta _{(\ell ),j}^{t+1})-\frac{1}{2} \} -m_0,\nonumber \\
\eta _{(\ell ),j}^{t+1}=\hspace{-0.15cm} \sum _{1\leq k\leq N}z_{jk}^{(\ell )}&x_k^{t+1}&+\hspace{-0.25cm} \sum _{1\leq k\leq N_{\ell -1}}\hspace{-0.25cm} u_{jk}^{(\ell )}x_k^{t}+\hspace{-0.15cm} \sum _{1\leq k\leq d}q_{jk}^{(\ell )}y_{k}^t-m_j^{(\ell )}.\ \ \label{backward_transition_model}
\end{eqnarray}
Here, the neurons in the $\ell $-th layer receive inputs from $\{ x_k^t\} _{1\leq k\leq N_{\ell -1}}$, $x^{t+1}$ and $y^t$ through the intermediate units, $\eta _{(\ell ),j}^{t+1}$, with the adjustable parameters $\kappa _{ij}$, $z_{jk}^{(\ell )}$, $u_{jk}^{(\ell )}$, $q_{jk} ^{(\ell )}$ and $m_j^{(\ell )}$ and the constant parameter $m_0$. This computation may seem strange, because here the neurons receive inputs from the future states. However, this is not problematic, because the goal of the computation is not to realize the states of the neural network but to calculate the value of $\log \nu (x^t|x^{t+1},y^t)$. Consider the following situation for this computation, for example. The intermediate units in the $\ell $-th layer receive inputs at time $t+1$ from $x^{t+1}$ and also from $\{ x_k^t\} _{1\leq k\leq N_{\ell -1}}$ and $y^t$ through some time-delay mechanisms. These intermediate units send outputs $\{ g(\eta _{(\ell ),j}^{t+1})\} _{1\leq j\leq M_{\ell }}$ to the neurons in the $\ell $-th layer. At this time, the $i$-th neuron in this layer possesses memory of its own state at time $t$, $x_i^t$, through some mechanism. Then, the $i$-th neuron can compute the value of $\nu _i(x_i^{t}|\{x_k^t\} _{k=1}^{N_{\ell -1}},x^{t+1}, y^t)$ as a function of $x_i^t$ and $\{ g(\eta _{(\ell ),j}^{t+1})\} _{1\leq j\leq M_{\ell }}$. The value of $\log \nu $ is the sum of these values of $\nu _i$ over the neurons in the neural network. \\
\\
{\it Proofs of the Relations Used in the Theoretical Analysis of the Reinforcement Learning Problem} : In this section, our goal is to prove Eq.(\ref{Bellman_optimal_solution}). First, we define the following functions called ``value functions'' in the field of reinforcement learning:
\begin{eqnarray}
\widehat{V} _{r, \alpha }^{(\gamma )}(y)&=&\mathrm{E} \left [\sum _{s=1}^{\infty }\gamma ^sr^{t+s}\Big{|} y^t=y\right ], \nonumber \\
\widehat{V} _{I, \alpha }^{(\gamma )}(y)&=&\mathrm{E} \left [\sum _{s=1}^{\infty }\gamma ^sI_{tr}^{t+s}\Big{|} y^t=y\right ], \nonumber \\
\widehat{V} _{\alpha }^{(\gamma )}(y)&=&\widehat{V}  _{I, \alpha }^{(\gamma )}(y)+\beta \widehat{V} _{r, \alpha }^{(\gamma )}(y).
\label{value_functions}
\end{eqnarray}
Then, we can write the learning problem Eq.(\ref{learning_problem}) in terms of the value functions as
\begin{eqnarray}
\mathrm{I} _{x\rightarrow y}+\beta \mathrm{E} [r^t]=\lim _{\gamma \rightarrow 1}(1-\gamma ) \widehat{V} _{\alpha }^{(\gamma )}(y),\ \ \ \ \forall y\in \mathcal{Y} . \label{learning_problem_rewritten}
\end{eqnarray}
By definition, the value function satisfies the following recursive relation called the ``Bellman equation'':
\begin{eqnarray}
\hspace{-0.5cm} \widehat{V} _{\alpha }^{(\gamma )}(y^{t})&=&\sum _{y^{t+1}\in \mathcal{Y} }\gamma \alpha (y^{t+1}|y^t)\nonumber \\
\times &\{ &\beta r(y^{t+1})-\log \alpha (y^{t+1}|y^t)+\widehat{V} _{\alpha }^{(\gamma )}(y^{t+1})\} . \label{Bellman_equation} 
\end{eqnarray}
Next, we show that for fixed $\gamma $, it is known that an optimal control $\alpha ^*(y^{t+1}|y^t)$ maximizes the value function at all $y \in \mathcal{Y} $, in comparison with suboptimal controls. Explicitly, for any control $\alpha $, the following inequality holds:
\begin{eqnarray}
\widehat{V} ^{(\gamma )}_{\alpha ^{*}}(y)\geq \widehat{V} ^{(\gamma )}_{\alpha }(y),\ \ \ \ \forall y\in \mathcal{Y} .\label{maximality_same_time}
\end{eqnarray}
\\
\\
In order to prove Eq.(\ref{maximality_same_time}), we consider the following operator called a backup operator, operating on functions of the environmental state $y^t$:
\begin{eqnarray}
B\phi (y^t)=\max _{\alpha (y^{t+1}|y^t)}\sum _{y^{t+1}\in \mathcal{Y} }\alpha (y^{t+1}|y^t)\{ \gamma r(y^{t+1})-\gamma \log \alpha (y^{t+1}|y^t)+\gamma \phi (y^{t+1})\} .
\end{eqnarray}
We first show that this operation results in contraction in the space of functions of environmental states $y^t$ with respect to max norm:
\begin{eqnarray}
\parallel \phi \parallel _{\infty }=\max _{y^t\in \mathcal{Y} }|\phi (y^t)|.
\end{eqnarray}
For two functions $\phi _1$ and $\phi _2$, a fixed $\alpha (y^{t+1}|y^t)$, and an operator $B_{\alpha }$ defined as
\begin{eqnarray}
B_{\alpha }\phi (y^t)=\sum _{y^{t+1}\in \mathcal{Y} }\alpha (y^{t+1}|y^t)\{ \gamma r(y^{t+1})-\gamma \log \alpha (y^{t+1}|y^t)+\gamma \phi (y^{t+1})\} ,
\end{eqnarray}
we have
\begin{eqnarray}
\parallel B_{\alpha }\phi _1-B_{\alpha }\phi _2 \parallel _{\infty }&=&\parallel \sum _{y^{t+1}}\alpha (y^{t+1}|y^t)\gamma \{ \phi _1(y^{t+1})-\phi _2(y^{t+1})\} \parallel _{\infty } \nonumber \\
&\leq &\max _{y^{t}\in \mathcal{Y} }\left \{ \sum _{y^{t+1}}|\alpha (y^{t+1}|y^t)|\right \} \parallel \gamma (\phi _1-\phi _2)\parallel _{\infty } \nonumber \\
&=&\gamma \parallel \phi _1-\phi _2\parallel _{\infty }.
\end{eqnarray}
Then, with the distribution $\alpha ^{(i),*}(y^{t+1}|y^{t})$ maximizing $B_{\alpha }\phi _i(y^t)$ $(i=1,2)$, we have 
\begin{eqnarray}
\parallel B\phi _1-B\phi _2\parallel _{\infty }&=&\parallel B_{\alpha ^{(1),*}}\phi _1-B_{\alpha ^{(2),*}}\phi _2\parallel _{\infty }\nonumber \\
&\leq &\max _i\parallel B_{\alpha ^{(i),*}}\phi _1-B_{\alpha ^{(i),*}}\phi _2\parallel _{\infty }\nonumber \\
&\leq &\gamma \parallel \phi _1-\phi _2\parallel _{\infty }.
\end{eqnarray}
This proves that the backup operation yields a contraction of the space of functions on the environmental state space $\mathcal{Y} $, and that there is a unique fixed point of this operation in this space of functions. Because the backup operation always increases the values of any value function at any point in $\mathcal{Y} $, we have Eq.(\ref{maximality_same_time}).
Hence, when we consider the maximality condition of $\widehat{V} _{\alpha }^{(\gamma )}(y)$ with respect to $\alpha (y^{t+1}|y^t)$, it is sufficient to consider the stationarity condition of $B\widehat{V} _{\alpha }^{(\gamma )}(y^t)$ by differentiating it with respect to $\alpha (y^{t+1}|y^t)$ and simply putting the derivative of $\widehat{V} _{\alpha }^{(\gamma )}(y^{t+1})$ to be zero. Solving the stationarity condition with the Lagrange multiplier corresponding to $\sum _{y^{t+1}}\alpha (y^{t+1}|y^t)=1$, we obtain 
\begin{eqnarray}
\alpha ^*(y^{t+1}|y^t)&\propto &\exp [\beta \{ r(y^{t+1})+\widehat{V} _{r,\alpha ^{*}}^{(\gamma )}(y^{t+1})\} +\widehat{V} _{I,\alpha ^{*}}^{(\gamma )}(y^{t+1})]. \label{optimal_solution_gamma}
\end{eqnarray} 
The optimal condition for the learning problem, the maximization of Eq.(\ref{learning_problem_rewritten}), is obtained by taking the limit $\gamma \rightarrow 1$ in Eq.(\ref{optimal_solution_gamma}). In order to avoid divergence, we need to replace the value functions in Eq.(\ref{optimal_solution_gamma}) with the functions defined in Eq.(\ref{excess_function}) that represent the ``excess reward'' and ``excess information''.
\\
\\
{\it Derivation and Biological Plausibility of the Learning Rule} : In the gradient ascent method used for the simulation, we update each parameter $\theta \in \{ \rho _{ij}, v_{jk}^{(\ell )}, w_{jk}^{(\ell )}, h _j^{(\ell )}, \kappa _{ij}, z_{jk}^{(\ell )}, u_{jk}^{(\ell )}, q_{jk} ^{(\ell )}, m_j^{(\ell )}\} $ as follows:
\begin{eqnarray}
\theta ^{t+1}&=&\theta ^t+\epsilon \left \{ \tau (\beta r^{t+1}-\Theta _{\nu }^t)\psi _{\theta }^t-\frac{\partial }{\partial \theta } \Theta _{\nu }^t\right \} \nonumber \\
\psi _{\theta }^t&=&\frac{1}{\tau } \frac{\partial }{\partial \theta } \log \pi (x^{t+1}|y^{t+1})+\left (1-\frac{1}{\tau }\right )\psi _{\theta }^{t-1}. \label{learning rule}
\end{eqnarray}
Here, the constant $\tau $ is a positive real number that is large compared with the mixing time of the dynamics. We set $\epsilon $ to such a small value that the change in the model parameters in each update does not affect the stationarity on a time scale of $\tau $. Then, in the above learning rule, the expectation value of the change in the parameter $\theta $ in each update is equal to the gradient of $\mathrm{E} [\beta r^t-\Theta _{\nu }^t] $ with respect to $\theta $ as we show below. Thus, we can regard the learning rule as a stochastic gradient ascent algorithm to maximize $\mathrm{E} [\beta r^t-\Theta _{\nu }^t] $.
\\
\\
In the gradient ascent method, we must calculate the gradient of the following quantity with respect to $\theta $:
\begin{eqnarray*}
\mathrm{E} [\beta r^t-\Theta _{\nu }^t]=\sum _{x^{t},x^{t+1},y^t,y^{t+1}}p_s(y^t)\pi (x^t|y^{t})\mu (y^{t+1}|y^t,x^t)\pi (x^{t+1}|y^{t+1})\{ \beta r(y^{t+1})-\Theta _{\nu }^{t}\} .
\end{eqnarray*}
In this calculation, we find that differentiation of the stationary distribution $p_s(y^t)$ is apparently intractable, while differentiation of the other components is easily carried out. We note, however, that we do not need to differentiate the stationary distribution explicitly, assuming that the stationary distribution is a smooth function of any model parameter $\theta $. In this case, small changes in $p_s(y^{t-\tau })$ for $\tau \gg 1$ vanish at $t$ and $t-1$, and thus terms including the derivatives of $p_s(y^{t-\tau })$ are negligible (see also \cite{Baxter99directgradient-based}). Thus, we can compute the gradient as follows: 
\begin{eqnarray}
&&\frac{\partial }{\partial \theta } \mathrm{E} [\beta r(y^{t+1})-\Theta _{\nu }^{t}]\nonumber \\
&=&\lim _{\tau \rightarrow \infty }\sum _{y^{t-\tau }, x^{t-\tau },\cdots ,y^{t+1},x^{t+1}}p_s(y^{t-\tau })\frac{\partial }{\partial \theta } \left [ \pi (x^{t-\tau }|y^{t-\tau })\prod _{s=0}^{\tau }\mu (y^{t-s+1}|x^{t-s},y^{t-s})\pi (x^{t-s+1}|y^{t-s+1}) \{ \beta r(y^{t+1})-\Theta _{\nu }^{t}\} \right ] \nonumber \\
&=&\lim _{\tau \rightarrow \infty }\sum _{y^{t-\tau }, x^{t-\tau },\cdots ,y^{t+1},x^{t+1}}p_s(y^{t-\tau }, x^{t-\tau },\cdots ,y^{t+1},x^{t+1})\sum _{s=0}^{\tau +1}\frac{\frac{\partial }{\partial \theta } \pi (x^{t-s+1}|y^{t-s+1})}{\pi (x^{t-s+1}|y^{t-s+1})} \{ \beta r(y^{t+1})-\Theta _{\nu }^{t}\} \nonumber \\ 
&\ &+\lim _{\tau \rightarrow \infty }\sum _{y^{\tau }, x^{t-\tau },\cdots ,y^{t+1},x^{t+1}}p_s(y^{t-\tau }, x^{t-\tau },\cdots ,y^{t+1},x^{t+1})\frac{\partial }{\partial \theta } \{ \beta r(y^{t+1})-\Theta _{\nu }^{t}\} \nonumber \\
&=&\lim _{\tau \rightarrow \infty }\mathrm{E} \left [\{ \beta r(y^{t+1})-\Theta _{\nu }^t \} \sum _{s=0}^{\tau +1}\frac{\partial }{\partial \theta } \log \pi (x^{t-s+1}|y^{t-s+1})\right ]-\mathrm{E} \left [\frac{\partial }{\partial \theta } \Theta _{\nu }^{t}\right ]. \label{derivative_batch}
\end{eqnarray} 
Note that the third equality follows from $\frac{\partial }{\partial \theta } r(y^{t+1})=0$. In order to decompose the expectation values into time-stepwise quantities, we introduce the auxiliary variable $\psi _{\theta }^t$, defined through
\begin{eqnarray}
\psi _{\theta }^t=\frac{1}{\tau } \frac{\partial }{\partial \theta } \log \pi (x^{t+1}|y^{t+1})+(1-\frac{1}{\tau } )\psi _{\theta }^{t-1},\ \ \ \mathrm{and} \ \ \ \psi _{\theta }^t=0 \ \ (t\leq 0). \label{online_process}
\end{eqnarray}
Then, we have
\begin{eqnarray*}
\psi _{\theta }^t=\frac{1}{\tau } \sum _{s=0}^{\infty }(1-\frac{1}{\tau } )^s\frac{\partial }{\partial \theta } \log \pi (x^{t-s+1}|y^{t-s+1}). 
\end{eqnarray*}
If the process under consideration is stationary, $\psi _{\theta }^t$ approaches the long-time average of $\frac{\partial }{\partial \theta } \log \pi (x^{t+1}|y^{t+1})$ as $\tau \rightarrow \infty $ and $t/\tau \rightarrow \infty $. Similarly, assuming that the correlation of $\beta r(y^{t+1})-\Theta _{\nu }^t$ with $\frac{\partial }{\partial \theta } \log \pi (x^{t-\tau +1}|y^{t-\tau +1})$ is small for $\tau \gg 1$ and that $T\gg \tau $, we have
\begin{eqnarray}
\mathrm{E} \left [\{ \beta r(y^{t+1})-\Theta _{\nu }^{t}\} \sum _{s=0}^{\tau +1}\frac{\partial }{\partial \theta } \log \pi (x^{t-s+1}|y^{t-s+1})\right ]&\approx &\frac{\tau }{T} \sum _{t=1}^{T}(\beta r(y^{t+1})-\Theta _{\nu }^t)\psi _{\theta }^t.\label{online_correlation}
\end{eqnarray}
Then, applying a well-known argument in stochastic approximation theory \citep{robbins1951stochastic}, we obtain the learning rule given in Eq.(\ref{learning rule}) as a stepwise approximation of the gradient in Eq.(\ref{derivative_batch}). \\
\\
Finally, we derive the exact form of the learning rule with respect to several $\theta $ and present its interpretation. Note that Eq.(\ref{learning rule}) is composed of $\log \pi (x^{t+1}|y^{t+1})$, $\log \nu (x^t|x^{t+1},y^t)$  and their derivatives with respect to $\theta $. First, we show that these components are easily calculated in a neuron-wise manner. Note that $\Theta _{\nu }^t$, $\log \pi (x^{t+1}|y^{t+1})$ and $\log \nu (x^t|x^{t+1},y^t)$ are decomposed as
\begin{eqnarray}
\Theta _{\nu }^t&=&\log \pi (x^{t+1}|y^{t+1})-\log \nu (x^t|x^{t+1},y^t) \nonumber \\
&=&\sum _{\ell =1}^L\sum _{i=N_{\ell -1}+1}^{N_{\ell }} \log \pi _i(x_i^{t+1}|y^{t+1}, \{ x_k^{t+1}\} _{k=1}^{N_{\ell -1}})-\sum _{\ell =1}^L\sum _{i=N_{\ell -1}+1}^{N_{\ell }}\log \nu _i(x_i^t|\{ x_k^t\} _{k=1}^{N_{\ell -1}}, x^{t+1}, y^t) \nonumber \\
&=&\sum _i\log \chi (e_i^{t+1}, x_i^{t+1})-\sum _i\log \chi (f_i^{t}, x_i^{t}),
\end{eqnarray}
where
\begin{eqnarray}
\chi (a,b)&=&\left \{ 
\begin{array}{ccc}
g(a)&,\ & \mathrm{if\ \ } b=1, \nonumber \\
-g(a)&,\ & \mathrm{if\ \ } b=0. \nonumber 
\end{array}
\right. \\
\end{eqnarray}
Then, the derivatives of $\Theta _{\nu }^t$, $\log \pi (x^{t+1}|y^{t+1})$ and $\log \nu (x^t|x^{t+1},y^t)$ (with respect to $v_{jk}^{(\ell )},\rho _{ij},z_{jk}^{(\ell )},\kappa _{ij}$, for example) are calculated as follows.
First, denoting the derivative of $\chi (a,b)$ with respect to $a$ as $\chi _a(a,b)$, 
\begin{eqnarray}
\frac{\partial }{\partial v_{jk}^{(\ell )}} \log \pi (x^{t+1}|y^{t+1})&=&\sum _{i=N_{\ell -1}+1}^{N_{\ell }} \frac{\partial }{\partial v_{jk}^{(\ell )}}\log \pi _i(x_i^{t+1}|y^{t+1},\{ x_k^{t+1}\} _{k=1}^{N_{\ell -1}} ) \nonumber \\
&=&\sum _{i=N_{\ell -1}+1}^{N_{\ell }}\frac{\chi _a(e_i^{t+1},x_i^{t+1})}{\chi (e_i^{t+1},x_i^{t+1})} \rho _{ij}g^{\prime }(\xi _{(\ell ),j}^{t+1})y_k^{t+1}. 
\end{eqnarray}
\begin{eqnarray}
\frac{\partial }{\partial \rho _{ij}} \log \pi (x^{t+1}|y^{t+1})&=&\frac{\partial }{\partial \rho _{ij}}\log \pi _i(x_i^{t+1}|y^{t+1},\{ x_k^{t+1}\} _{k=1}^{N_{\ell -1}} ) \nonumber \\
&=&\frac{\chi _a(e_i^{t+1},x_i^{t+1})}{\chi (e_i^{t+1},x_i^{t+1})} g(\xi _{(\ell ,j)}^{t+1}). 
\end{eqnarray}
\begin{eqnarray}
\frac{\partial }{\partial z_{jk}^{(\ell )}} \log \nu (x^t|x^{t+1},y^t)&=&\sum _{i=N_{\ell -1}+1}^{N_{\ell }} \frac{\partial }{\partial z_{jk}^{(\ell )}}\log \nu _i(x_i^{t}|\{ x_k^{t}\} _{k=1}^{N_{\ell -1}},x^{t+1},y^{t}) \nonumber \\
&=&\sum _{i=N_{\ell -1}+1}^{N_{\ell }}\frac{\chi _a(f_i^{t+1},x_i^{t})}{\chi (f_i^{t+1},x_i^{t})} \kappa _{ij}^{(\ell )}g^{\prime }(\eta _{(\ell ),j}^{t+1})x_k^{t+1}. 
\end{eqnarray}
\begin{eqnarray}
\frac{\partial }{\partial \kappa _{ij}} \log \nu (x^t|x^{t+1},y^t)&=&\frac{\partial }{\partial \kappa _{ij}}\log \nu _i(x_i^{t+1}|\{ x_k^{t}\} _{k=1}^{N_{\ell -1}},x^{t+1},y^{t}) \nonumber \\
&=&\frac{\chi _a(f_i^{t+1},x_i^{t})}{\chi (f_i^{t+1},x_i^{t})} g(\eta _{(\ell ,j)}^{t+1}). 
\end{eqnarray}
It should be noted that calculations of the derivatives involve quantities only for related neurons and intermediate units. For example, the derivative with respect to $v_{jk}^{(\ell )}$ used only information regarding $y_k^{t+1}$, $\xi _{(\ell ),k}^{t+1}$ and $\{ \mu _i^{t+1}, x_i^{t+1}, \rho _{ij}\} $ of the $i$-th neuron to which the $j$-th intermediate unit is connected. Thus, we can regard the change in the synaptic strength $v_{jk}^{(\ell )}$ as being determined by the local interactions at the synapse on the $j$-th intermediate unit. Continuing with this line of argument, we can obtain even more realistic forms of learning rules for actual neural systems. However, we do not go into detail here, because the argument becomes quite complicated and is beyond the scope of the current study.
\\
\\
{\it Initial Values of Model Parameters and Values of Learning Parameters Used in the Simulation } : In the numerical simulation of our model of learning, we used initial values of the model parameters that results in behavior in which the animal primarily attempts to avoid negative reward, mimicking innate behavior of real animals. We set the values of the model parameters involved in the inputs to the movement-related neurons as shown in Fig.6. A neuron controlling motion in one of four directions receives connections with relatively strong positive weights, $\rho _0$, from a specialized intermediate unit (for example, from $\xi _{(4),1}^t$ to $x_N^t$). The intermediate unit receives connections from the environmental variables that take the values of the rewards within one step of the animal's position, $y_k^t$ ($4\leq k\leq 12$), with the weight-values $v_0$, $-v_0$ and $0$, as illustrated in Fig.6. These initial values of the weight parameters make the neurons controlling motion take a value of 1 when relative amounts of the reward in the corresponding direction are large. We chose the other weight parameters with small random values in accordance with the following:
$\rho _{ij}\sim [-0.05:0.05]\ (i\leq N_2)$; $v_{ij}^{(\ell )}\sim [-0.05:0.05]\  (\ell =1,2)$; $\rho _{ij}=0$ if $N_2< i\leq N_4=N$ and $(i, j)\neq (N,1),(N-1,2),(N-2,1),(N-3,2)$; $v_{ij}^{(\ell )}=0$ ($\ell =3,4$ except the red and blue synaptic weights in Fig.6); $w_{ij}^{(\ell )}\sim [-0.05:0.05]\  (\ell =1,2)$; $w_{ij}^{(\ell )}=0\ (\ell =3,4)$; $h_0=\log 20$; $h_j^{(\ell )}=0$; $\kappa _{ij}\sim [-0.05:0.05]$; $z_{jk}^{(\ell )}\sim [-0.05:0.05]$; $u_{jk}^{(\ell )}\sim [-0.05:0.05]$; $q_{jk}^{(\ell )}\sim [-0.05:0.05]$; $m_j^{(\ell )}=0$; $m_0=0$.\\
In the updates of the model parameters according to Eq.(\ref{learning rule}), we used the following (fixed) values of learning parameters: $\epsilon =3.0\times 10^{-5}$; $\tau =50$. 
\begin{figure}
\includegraphics[width=85mm]{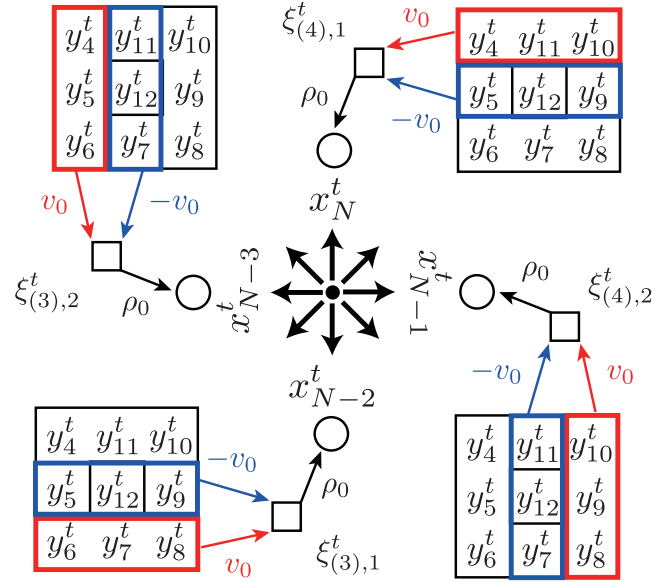}
\caption{(Color online). Initial values of model parameters for synaptic weights to the movement-related neurons.}
\end{figure}
\end{widetext}
\bibliography{ref.bib}

\end{document}